\documentclass{PoS}

\usepackage{subcaption}
\usepackage{amsmath,amssymb}

\title{Determination and application of TMD parton densities using the Parton Branching method}

\ShortTitle{TMDs with Parton Branching method}

\author{\speaker{A. Bermudez Martinez}$^{1}$, P. Connor$^{1}$, F. Hautmann$^{2,3,4}$, H. Jung$^{1}$, A. Lelek$^{1a}$,}
\author{V. Radescu$^{3,5b}$, R. \v{Z}leb\v{c}\'{i}k$^{1}$\\
        $^{1}$DESY, Hamburg, FRG\\
        $^{2}$Rutherford Appleton Laboratory, Chilton OX11 0QX\\
        $^{3}$University of Oxford, Oxford OX1 3NP\\
        $^{4}$Elementary Particle Physics, University of Antwerp, B 2020 Antwerp\\
        $^{5}$CERN, CH-1211 Geneva 23\\
        $^{a}$Now at University of Antwerp\\
        $^{b}$Now at IBM Germany\\
        E-mail: \email{armando.bermudez.martinez@desy.de}}

\abstract{We present a determination of parton densities at NLO obtained with the Parton Branching method using precision measurements of deep inelastic scattering cross sections at HERA. The two sets of parton densities shown in this work are obtained with the same angular angular ordering condition for the evolution scale and they differ in the chosen scale for the $\alpha_s$ evaluation, for which we consider two scenarios: the evolution scale, and the transverse momentum $\rm{q}_T$ from the angular ordering prescription. The transverse momentum dependent densities obtained with the Parton Branching method are applied to two LHC processes: the Drell-Yan $\rm{p}_T$ spectrum and the azimuthal correlation in high $\rm{p}_T$ dijet events. For the Drell-Yan $\rm{p}_T$ spectrum a significant effect from the $\alpha_s$ scale choice is observed.}

\FullConference{XXVI International Workshop on Deep-Inelastic Scattering and Related Subjects (DIS2018)\\
		16-20 April 2018\\
		University of Kobe, Japan}

\begin{document}

\section{Introduction}
\label{sec:intro}

The Parton Branching (PB) method, introduced in Refs~\cite{Hautmann:2017xtx,Hautmann:2017fcj,Martinez:2018jxt}, and implemented in the xFitter package~\cite{Alekhin:2014irh}, provides an iterative solution for the evolution of both collinear and transverse momentum dependent parton distributions (TMDs)~\cite{Angeles-Martinez:2015sea}. 
An important feature of this method is that it gives a solution which is fully exclusive, meaning that the splitting kinematics at each branching vertex is known. Therefore it allows for a natural determination of the transverse momentum TMDs, as the transverse momentum at every branching vertex is known. The PB method has shown to be valid for leading-order (LO), next-to-LO (NLO) and next-to-NLO (NNLO). Along this line, the agreement within a percent of the integrated TMDs obtained by this method, with the corresponding semi-analytical solution of the DGLAP~\cite{Gribov:1972ri,Dokshitzer:1977sg,Altarelli:1977zs} evolution (implemented in Qcdnum~\cite{Botje:2010ay}) was achieved at LO, NLO and NNLO~\cite{Hautmann:2017xtx,Hautmann:2017fcj,Radek:2017pos}. In this report the determination of parton densities at NLO obtained using the PB method is presented together with applications of the obtained TMDs to LHC processes (Drell-Yan and azimuthal correlations in dijet events). 

\section{Parton Branching method}
\label{sec:pb} 

The evolution with the scale $Q^2$ of a parton density $f_a(x,Q^2)$ of flavour $a$ and longitudinal momentum fraction $x$ can be described by the DGLAP equation. The PB method, applied to a momentum weighted parton density $\hat{f}_a(x,Q^2)=xf_a(x,Q^2)$, is based on the following equation~\cite{Hautmann:2017fcj}:

\small
\begin{equation}
Q^2\dfrac{\partial \hat{f}_a(x,Q^2)}{\partial Q^2} =  \sum_b \left[\int_x^{z_{\text{max}}} dz P^R_{ab}\left(z,\alpha_s(Q^2)\right) \hat{f}_b(\dfrac{x}{z},Q^2) - \hat{f}_a(x,Q^2)\int_0^{z_{\text{max}}} dz\ z P^R_{ba}\left(z,\alpha_s(Q^2)\right)\right] ,
\label{eq1}
\end{equation} 
\normalsize
where $z$ is the light-cone momentum fraction of the parton undergoing the splitting, and $P^R_{ab}\left(z,\alpha_s(Q^2)\right)$ are the real DGLAP splitting kernels~\cite{Hautmann:2017fcj,Martinez:2018jxt}. The parameter $z_{\text{max}}$ is the resolution scale~\cite{Hautmann:2017xtx} that separates the resolvable emissions region ($z < z_{\text{max}}$) from the non-resolvable emissions one ($z > z_{\text{max}}$). The Eq.~\ref{eq1} converges to the DGLAP equation in the limit $z_{\text{max}} \to 1$~\cite{Hautmann:2017xtx}. The indexes $a$ and $b$ denote the flavor indexes. The Eq.~\ref{eq1} can then be rewritten as:

\small
\begin{equation}
Q^2\dfrac{\partial}{\partial Q^2}\dfrac{\hat{f}_a(x,Q^2)}{\Delta_a(Q^2,Q_0^2)} = \dfrac{1}{\Delta_a(Q^2,Q_0^2)} \sum_b\int_x^{z_{\text{max}}} dz  P^R_{ab}\left(z,\alpha_s(Q^2)\right) \hat{f}_b(\dfrac{x}{z},Q^2) ,
\label{eq2}
\end{equation}
\normalsize
where the Sudakov form factor was introduced, and it is defined as:

\small
\begin{equation}
\Delta_a(Q^{2},Q^{2}_0) \equiv \exp\left[- \sum_b \int_{Q_0^2}^{Q^2}\frac{dQ^{'2}}{Q^{'2}} \int_0^{z_{\text{max}}}dz\ z P^R_{ba}\left(z,\alpha_s(Q^{'2})\right)\right] 
\label{eq3}
\end{equation}
\normalsize

The sensitivity of Eq.~\ref{eq1} to the arbitrary parameter $z_{\text{max}}$ has been studied in Ref.~\cite{Hautmann:2017fcj} and shown to give stable solutions in the limit of applicability ($1 - z_{\text{max}} \ll 1$).

The Sudakov form factor $\Delta_a(Q^{2},Q^{2}_0)$ represents the no-branching probability between the scales $Q^{2}_0$ and $Q^{2}$ for a parton with flavor $a$. 
The probability of no-branching between two scales $Q^{2}_1$, $Q^{2}_2$ can be expressed as $\Delta_a(Q^{2}_2,Q^{2}_1)=\Delta_a(Q^{2}_2,Q^{2}_0)/\Delta_a(Q^{2}_1,Q^{2}_0)$. The DGLAP evolution is then explicitly obtained by a sequence of branchings happening at scales $Q^{2}_0<Q^{2}_1<\hdots<Q^{2}$. The scale $Q^{2}_i$ of the $i$-th splitting is chosen (via Monte Carlo (MC) sampling) according to the distribution $\Delta_a(Q^{2}_i,Q^{2}_{i-1})$ which depends on the previous splitting. Analogously, the corresponding $z$ of the $i$-th splitting $a \to bc$ is generated according $P^R_{ba}(z,Q^{2}_i)$. The details of the procedure can be found in Ref.~\cite{Hautmann:2017fcj}.   

\section{Mapping the branching kinematics to the evolution of the TMDs}
\label{sec:ev}

Due to the branchings in the evolution, a parton of flavor $a$ and longitudinal momentum fraction $x$ acquires a transverse momentum $\mathbf{k_t}$. Consequently, the corresponding TMD $A_a(x,\mathbf{k_t};Q^2)$ evaluated at the scale $Q^2$ will depend also on the $\mathbf{k_t}$. The evolution of a TMD $A(x,\mathbf{k_t};Q^2)$ from a scale $Q^{2}_0$ to a scale $Q^{2}$ obeys the evolution equation introduced in~\cite{Hautmann:2017fcj,Martinez:2018jxt}:

\small
\begin{multline}
A_a(x,\mathbf{k_t};Q^2)= \Delta_a(Q^{2},Q^{2}_0)A_a(x,\mathbf{k_t};Q^2_0)+ \\
+ \sum_b \int_{Q_0^2}^{Q^2}\frac{d^2\mathbf{Q^{'}}}{\pi Q^{'2}}\dfrac{\Delta_a(Q^{2},Q^{2}_0)}{\Delta_a(Q^{'2},Q^{2}_0)} \int_x^{z_{\text{max}}} dz P^R_{ab}\left(z,\alpha_s(Q^{'2})\right) A_b\left(\dfrac{x}{z},\mathbf{k_t}+(1-z)\mathbf{Q'};Q^{'2}\right) .
\label{eq4}
\end{multline}
\normalsize

In the application of Eq.~\ref{eq4} we consider the scale at which $\alpha_s$ is evaluated not necessarily equal to the evolution scale~\cite{Martinez:2018jxt}. 
The kinematical variables associated with the splitting have to be related to the scales at which the TMD and the $\alpha_s$ are evaluated respectively. 
In this work we use the choice $Q=\rm{q}_T/(1-z)$ where $\rm{q}_T$ stands for the 
transverse momentum
of the emitted parton with respect to the beam axis. This choice enforces an angular ordering of the emissions therefore ensuring quantum coherence of softly radiated partons.

In the case of $\alpha_s$ two choices for the renormalization scale $Q^2_r$ are investigated in this work: the condition $Q_r=Q$ and the condition $Q_r=\rm{q}_T$ from the angular ordering prescription. 
  
\section{Determination of TMDs from fits to inclusive measurements}
\label{sec:fit}

\begin{figure}
\begin{subfigure}{.5\textwidth}
  \centering
  \includegraphics[width=1.\textwidth]{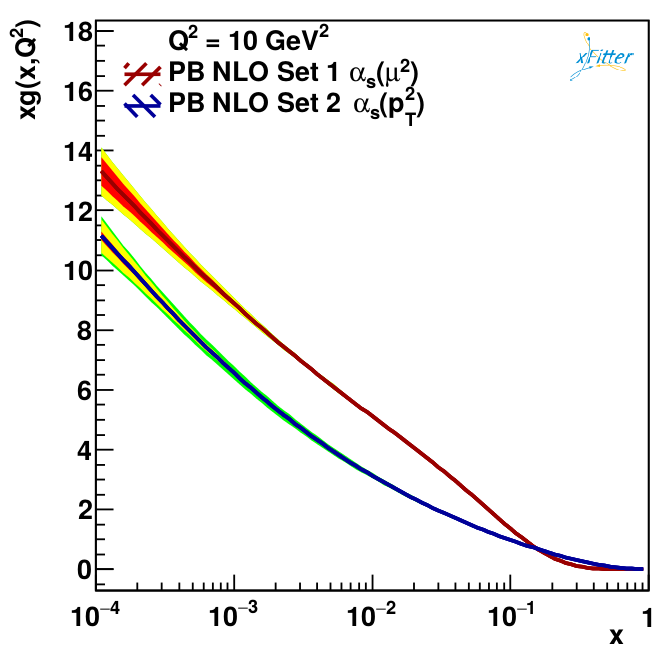}
  \caption{} 
  \label{fig:sfig1_1}
\end{subfigure}%
\begin{subfigure}{.5\textwidth}
  \centering
  \includegraphics[width=1.\textwidth]{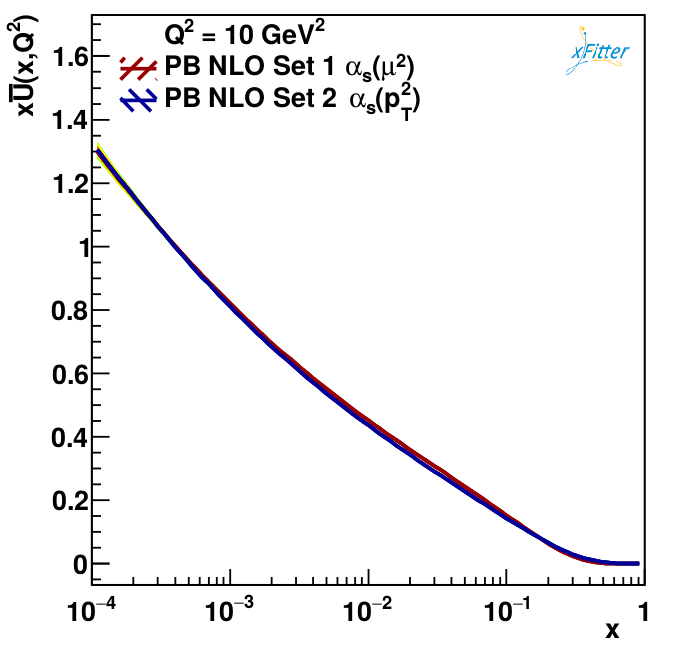}
  \caption{}
  \label{fig:sfig1_2}
\end{subfigure}
\caption{
Parton densities for different choices for the renormalization scale in $\alpha_s$ are shown. The red band shows the experimental uncertainty, the yellow
band the model dependence. The green band shows the uncertainty from the variation of the cut used to avoid the non-perturbative region in the $\alpha_s$ evaluation~\cite{Martinez:2018jxt}.}
\label{fig:fig1}
\end{figure}

Two TMD sets (Set 1 and Set 2) were obtained in Ref.~\cite{Martinez:2018jxt} using the two different choices $Q_r=Q$ and $Q_r=\rm{q}_T$ for the $\alpha_s$ evaluation (Fig.~\ref{fig:fig1}). For both sets the angular ordering condition $Q=\rm{q}_T/(1-z)$ was chosen for the evolution scale. The Fig.~\ref{fig:sfig1_1} shows that the sets differ in the gluon component at small scales.

The integrated TMD is fitted, within the xFitter where the PB method is implemented, to precision measurements of neutral and charged current DIS at various beam energies from HERA 1+2~\cite{Abramowicz:2015mha} in the ranges $3.5 < Q^2 < 50000$ GeV$^2$ and $4 \cdot 10^{-5} < x < 0.65$. The fits of both sets resulted in a similar $\chi^2/ndf\approx1.21$~\cite{Martinez:2018jxt}.

\section{TMDs applied to Drell-Yan and dijet azimuthal correlations at the LHC}
\label{sec:app}

\begin{figure}
\begin{subfigure}{.5\textwidth}
  \centering
  \includegraphics[width=1.\textwidth]{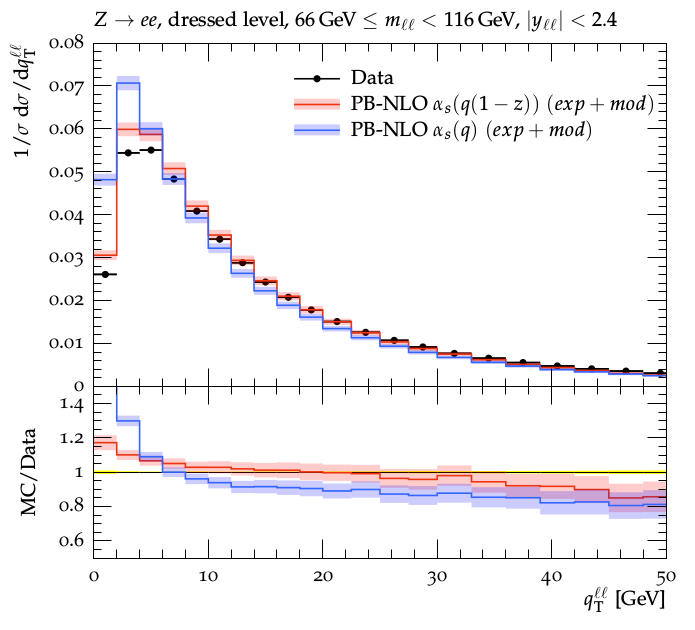}
  \caption{}
  \label{fig:sfig2_1}
\end{subfigure}%
\begin{subfigure}{.5\textwidth}
  \centering
  \includegraphics[width=1.\textwidth]{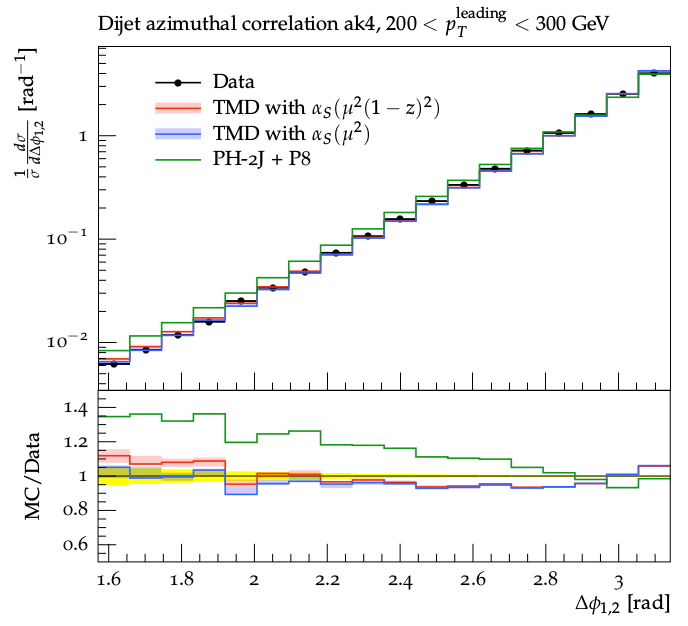}
  \caption{}
  \label{fig:sfig2_2}
\end{subfigure}
\caption{
a) The transverse momentum spectrum of the di-lepton pair measured by ATLAS~\cite{Aad:2015auj} and b) the $\Delta\phi_{12}$ in dijet events measured by CMS~\cite{Sirunyan:2017jnl} compared with the predictions based
on Set 1 and Set 2 TMD densities.}
\label{fig:fig2}
\end{figure}

In Fig.~\ref{fig:sfig2_1} the predictions for the transverse momentum spectrum of the Z-boson obtained with the two TMD distributions, and compared with the measurement from ATLAS~\cite{Aad:2015auj} are shown. The details of the calculation can be found in Ref.~\cite{Martinez:2018jxt}. Although the overall shape of the distribution is described by both TMD sets, the Set 2, for which $\alpha_s$ is evaluated at the scale $\rm{q}_T$ from the angular ordering prescription, provides a better description of the data. The transverse momentum spectrum of the Z-boson can therefore be used to investigate the higher order effects covered by the different scale choices. The reader should notice that no adjustment has been made and only by means of the $k_t$ from the TMDs (which are constrained by the data) we are able to reproduce the Z-boson transverse momentum spectrum up to a level of 20\%.  

Figure~\ref{fig:sfig2_2} illustrates the predictions for the azimuthal angular separation between the two leading jets ($\Delta\phi_{12}$) in inclusive dijet events, compared with the measurement from CMS~\cite{Sirunyan:2017jnl}. The events obtained with POWHEG~\cite{Alioli:2008gx} are convoluted with the TMDs and then showered. The CASCADE MC generator~\cite{Jung:2010si,Bury:2017jxo} was used for generating the initial state shower according to the TMD evolution whereas the PYTHIA6~\cite{Sjostrand:2006za} was used for the final state shower. The prediction from POWHEG + PYTHIA8~\cite{Sjostrand:2007gs} is also shown in Fig.~\ref{fig:sfig2_2}.

\section{Conclusions}
\label{sec:con}

The PB method has been utilized to obtain TMDs in a novel way. The integrated TMDs have been fitted to DIS data measured at the HERA experiment over a large range in $x$ and $Q^2$. Two TMD sets differing in the renormalization scale at which the $\alpha_s$ is evaluated were obtained and applied to LHC processes (Drell-Yan and azimuthal correlations in dijet events). The fits gave similar $\chi^2/ndf\approx1.21$ values for both sets. The Drell-Yan spectrum (Fig.~\ref{fig:sfig2_1}) is sensitive to the different gluon contributions in the TMDs. It is better described when $\alpha_s$ is evaluated at the scale $\rm{q}_T$ from the angular ordering prescription. On the other hand, for the azimuthal correlation between the leading jets in dijet events at LHC, both TMD variants give equivalent results.

\bibliography{skeleton}
\bibliographystyle{hieeetr}






\end{document}